\documentclass[twocolumn,showpacs,preprintnumbers,amsmath,amssymb]{revtex4}
\usepackage{graphicx}
\usepackage{epstopdf}
\usepackage{color}

\begin{document}

\preprint{APS/123-QED}

\title{Direct comparison of domain wall behavior in Permalloy nanowires patterned by electron beam lithography and focused ion beam milling}% Force line breaks with \\

\author{M.A. Basith*}
\affiliation{
School of Physics \& Astronomy, University of Glasgow, G12 8QQ, United Kingdom.
}

\author{S. McVitie}
\affiliation{
School of Physics \& Astronomy, University of Glasgow, G12 8QQ, United Kingdom.
}

\author{D. McGrouther}
\affiliation{
School of Physics \& Astronomy, University of Glasgow, G12 8QQ, United Kingdom.
}

\author{J. N. Chapman}
\affiliation{
School of Physics \& Astronomy, University of Glasgow, G12 8QQ, United Kingdom.
}
\author{J.M.R. Weaver}
\affiliation{
School of Electronics and Electrical Engineering, University of Glasgow, Glasgow G12 8QQ, United Kingdom.
}


\begin{abstract}
Nominally identical permalloy nanowires, with widths down to 150 nm, were fabricated onto a single electron transparent Si$_{3}$N$_{4}$ membrane using electron beam lithography (EBL) and focused ion beam (FIB) milling. Transmission electron microscopy (TEM) experiments were performed to compare the nanostructures produced by these two techniques in what we believe is the first direct comparison of fabrication techniques for nominally identical nanowires. Both EBL and FIB methods produced high quality structures with edge roughness being of the order of the mean grain size 5 -10 nm observed in the continuous films. However, significant grain growth was observed along the edges of the FIB patterned nanowires. Lorentz TEM \emph{in situ} imaging was carried out to compare the magnetic behavior of the domain walls in the patterned nanowires with anti-notches present to pin domain walls. The overall process of domain wall pinning and depinning at the anti-notches showed consistent behaviour between nanowires fabricated by the two methods with the FIB structures having slightly lower characteristic fields compared to the EBL wires. However, a significant difference was observed in the formation of a vortex structure inside the anti-notches of the EBL nanowires after depinning of the domain walls. No vortex structure was seen inside the anti-notches of the FIB patterned nanowires. Results from micromagnetic simulations suggest that the vortex structure inside the anti-notch can be suppressed if the saturation magnetization (M$_{s}$) is reduced along the nanowires edges. A reduction of M$_{s}$ along the wire edges may also be responsible for a decrease in the domain wall depinning fields. Whilst the two fabrication methods show that well defined structures can be produced for the dimensions considered here, the differences in the magnetic behavior for nominally identical structures may be an issue if such structures are to be used as conduits for domain walls in potential memory and logic applications. \\

\end{abstract}

\maketitle
\section{Introduction} \label{I}
Understanding and controlling magnetic domain wall (DW) behavior in ferromagnetic nanowires is of fundamental scientific interest and important for their potential applicability in future spintronic devices such as magnetic logic \cite{ref1} and race-track memory \cite{ref2}. Micromagnetic simulations \cite{ref3,ref4,ref5}, magneto-optic Kerr effect magnetometry \cite{ref6,ref7}, off-axis electron holography \cite{ref8} and magnetic imaging techniques \cite{ref9,ref10} have contributed useful information for a greater understanding of it properties. To ensure the reliable operation of devices, performance variability must be reduced and high quality nanofabrication is extremely important. Previous reports \cite{ref11,ref12} indicate  that the structural roughness at the edges of the  nanostructures produced by fabrication processes affect the magnetic properties of the nanostructures. Amongst a variety of nanofabrication techniques, electron beam lithography (EBL) is widely used for high resolution submicron scale patterning of magnetic materials. This technique requires a number of steps including resist spinning, pattern exposure, metallization, removal of resist from the sample surface and lift off of the residual materials. In contrast, focused ion beam (FIB) based fabrication by milling is essentially a one step patterning process of a continuous film and an excellent tool for rapid device proto-typing \cite{ref9}. Nevertheless, FIB has some disadvantages \cite{ref13} especially due to the heavy Ga$^+$ ions used for the milling process. These include radiation induced damage and ion implantation. The extent to which the magnetic properties of nanostructures are affected by the differences in physical properties produced by patterning structures using EBL and FIB techniques has not been studied extensively.  Therefore, in the present investigation, nominally identical permalloy nanowires fabricated by these two nanofabrication techniques are compared through characterization of their physical nanostructure and magnetic behavior.\\

\section{Experimental details} \label{II}
\begin{figure}[hh]
\centering
\includegraphics[width=7.5cm]{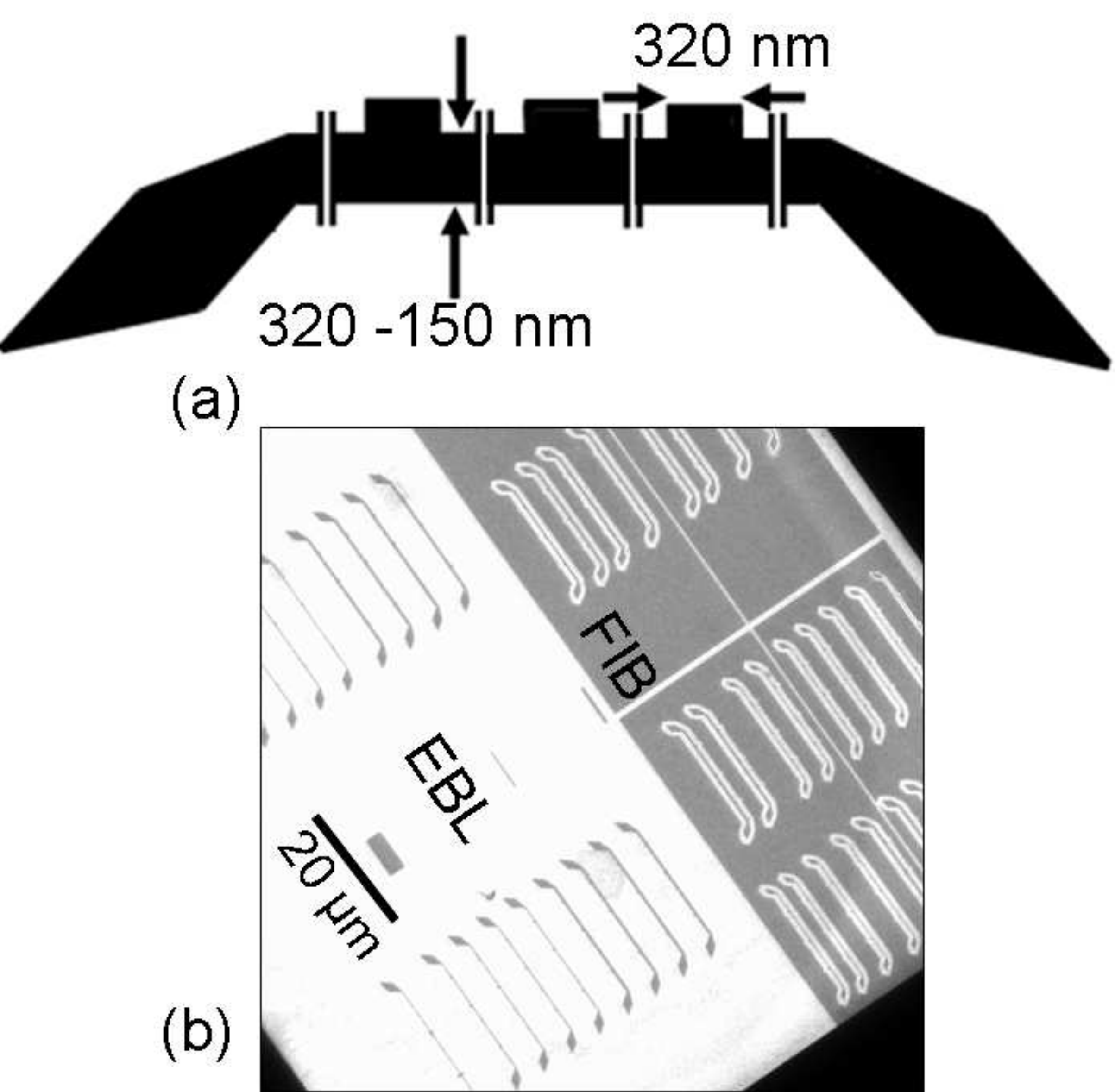}
\caption{(a) Schematic of the nanowire geometry. The distance between the anti-notches has been reduced for this schematic indicated by vertical bars. (b) TEM image shows the nanostructures written by EBL and FIB techniques. } \label{fig1}
\end{figure}

The permalloy (Ni$_{80}$Fe$_{20}$) film of 20 nm thickness was deposited by thermal evaporation onto a single electron transparent Si$_{3}$N$_{4}$ membrane for the fabrication of the nanowires. The structure of the fabricated nanowires is shown in Fig. \ref{fig1}(a) which contains three 320 nm wide and 140 nm high  rectangular protrusions referred to as anti-notches. Two wire widths were chosen for study, 320 and 150 nm, and both ends of the wire were connected to diamond shaped pads to allow controlled over the formation of DWs.  Nanostructures were fabricated first by an EBL/lift off process onto one half of the membrane. During the fabrication of the nanostructures using EBL/lift off technique, four large area rectangles were also fabricated onto the other half of this membrane. Identical structures were written inside these large rectangles containing a same film of permalloy (Py), as shown in Fig. \ref{fig1}(b), using FIB. This was done to ensure that the same film thickness and quality of the deposited Py was used for patterning using both the techniques.

For FIB patterning, a FEI Nova NanoLab 200 SEM/FIB workstation with a 10 nm full-width-half maximum Ga$^+$ ion probe (energy of 30 keV and current of 10 pA) was used. The FIB pattern generation involved a two-step vector milling strategy described elsewhere \cite{ref9}. Cross-sectional (x-sectional) TEM samples of the patterned nanowires were prepared by FIB milling to observe the details of the  physical structure of the nanowires around their edges. The preparation of the x-sectional TEM samples of thin films is always challenging, however, the fragility of the $Si_3N_4$ membrane makes it even more so. Nevertheless, attempts were made to make the TEM sample as thin as possible which is desirable for good imaging.  To obtain a thinner sample and to provide additional support during milling process prior to load the sample inside the FIB chamber, approximately 200 nm Al was deposited in to the back side of the membrane.  X-sectional TEM samples were prepared using the FIB based \emph{in situ} lift-out technique which is described in reference \cite{ref13}. Conventional TEM was used to investigate the physical structure while the Fresnel mode of Lorentz TEM (LTEM) was used to image the magnetic behavior of DWs in the patterned nanowires. LTEM imaging was performed in a Philips CM20 field emission gun TEM equipped with Lorentz lenses and suitable for performing \emph{in situ} magnetizing experiments \cite{ref14}.   \\

\section{Results and discussions} \label{III}
\subsection{Structural characterizations} \label{I}
\begin{figure}[!hh]
\centering
\includegraphics[width=7cm]{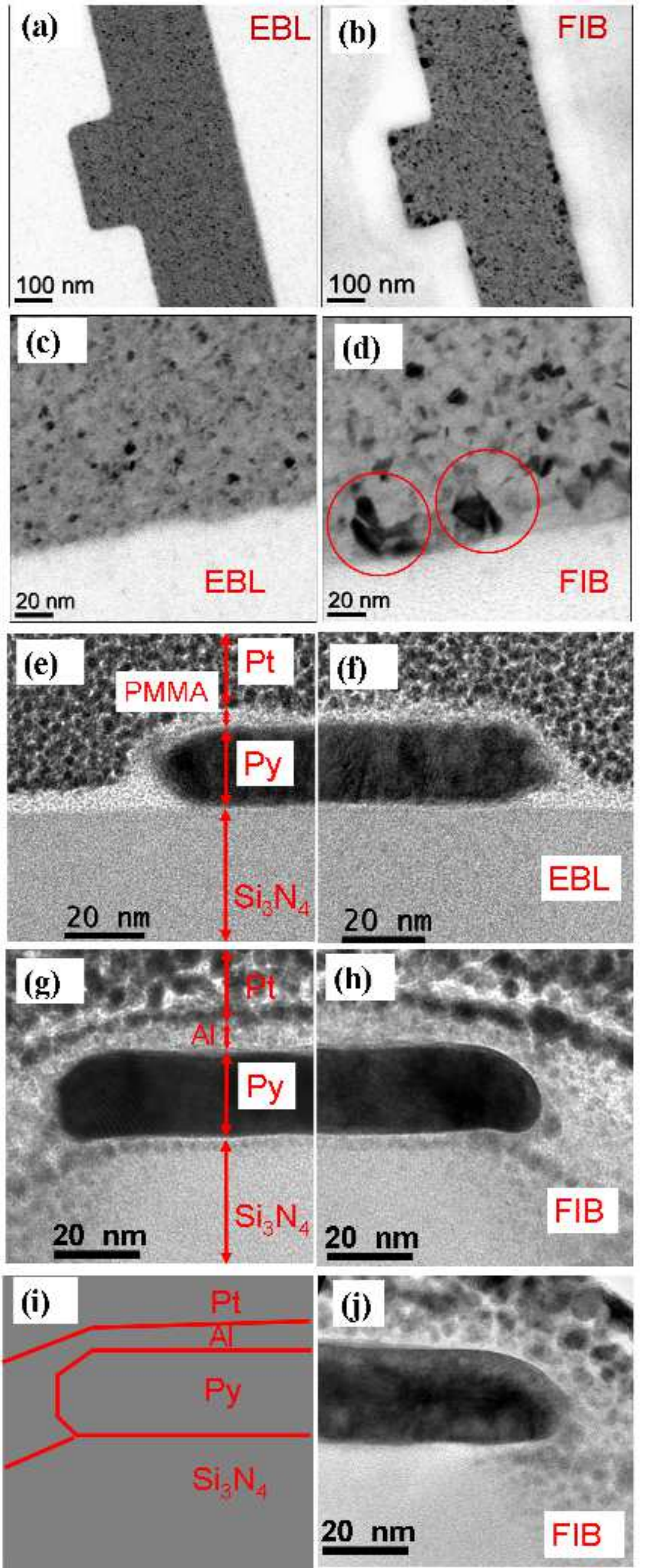}
\caption{(Color online) TEM bright field images of the plan view of 320 nm wide nanowires patterned by EBL (a,c) and FIB (b,d). Circle marked regions indicate grain growth along FIB patterned nanowire edge. X-sectional TEM bright field images showing the wire edges of the 320 nm wide nanowires patterned by EBL (e,f) and FIB (g,h), respectively. The individual layers are labeled in (e) and (g). Furthermore figure (i) shows schematically the individual layers of image (g). X-sectional image (j) is showing the one edge of a 150 nm wide FIB patterned nanowire. } \label{fig2}
\end{figure}

Figures \ref{fig2}(a-d) show  TEM bright field (BF) images of the plan view of 320 nm wide nanowires patterned by EBL (a,c) and FIB (b,d). These images reveal polycrystalline films with well defined anti-notches (Figs. \ref{fig2}(a, b)) produced by both methods. The measured edge roughness was of the order of the mean grain size 5 -10 nm observed in the continuous film. The grain structure of the EBL nanowires, as shown in Fig. \ref{fig2}(c), reveal a log-normal distribution \cite{ref15} of the grain size peaked around this mean value. For wires fabricated by FIB, significant grain growth was observed up to a distance of 30 to 40 nm from the wire edges, as marked by circles in Fig. \ref{fig2}(d). The size of these larger grains varied from 20 to 30 nm. However, the grain size distribution in the center of the FIB patterned wires is similar to the EBL/continuous film. Therefore, the differences between the physical nanostructures of the patterned nanowires are clearly associated with the regions close to the edge of the wires.  In a related study this effect was observed for nanoelements with a width of $<$ 100 nm where the grain growth was seen in the whole of the structure and actually was seen to be amorphous when the width was less than 60 nm \cite{ref16}. In that study the proximity of the edges meant that there was effectively no non-irradiated region in the centre of the structure unlike in this study.\\

X-sectional samples of the patterned nanowires were prepared by FIB lift-out technique and bright field TEM images were recorded to observe the morphology of the nanowires, particularly around their edges. Figures \ref{fig2}(e,f) show TEM bright field images of the wire edges of a 320 nm wide EBL nanowire. The individual layers are labeled in image Fig. \ref{fig2}(e). In the x-sectional images of the EBL sample, Figs. \ref{fig2}(e,f) the top layer is an electron beam deposited platinum (Pt) protective layer. The white band around the edge structure is residual PMMA and its thickness is approximately 3-5 nm which is consistent with the measured residual PMMA thickness after acetone removal in lift-off \cite{ref17}. Furthermore these images reveal a tapered edge structure of the patterned nanowire. Formation of the tapered edge structure in this case may be due to shadowing associated with a build-up of metal during evaporation on the residual PMMA resist to the side of the nanowire. By comparison x-sectional TEM images,  Figs. \ref{fig2}(g,h) show the wire edges of a 320 nm wide FIB milled nanowire. Figure \ref{fig2}(i) is a schematic showing the individual layers of \ref{fig2}(g). To distinguish between the Py layer and Pt over layer, about 5 nm of Al was deposited on top of the Py prior to Pt deposition. In the case of the EBL patterned nanowire, the residual PMMA meant that such an intermediate layer was unnecessary. A comparison of the edge structure from the x-sectional images of the EBL and FIB indicate that in both cases the edge profile is not vertical but tapered, though in the case of the FIB structure the tapering is slightly less symmetric. Whilst in the case of the EBL nanowires the tapering may be in part due to residual resist, in the case of the FIB nanowires this is clearly due to the interaction of the Ga$^+$ ion beam with the film during the milling process. However in both cases the edge structure can be described as well defined.  A x-sectional TEM image from a FIB milled 150 nm wide wire, Fig. \ref{fig2}(j), also demonstrates a well defined edge profile of this narrower wire consistent with that observed in the wider nanowires.

From the plan view and cross-sectional TEM bright field images, Fig. \ref{fig2}     it is apparent that the main difference between the wires fabricated by the two processes is the grain structure observed along the edges of the nanowires, in particular the formation of larger grains along edges of the FIB milling nanowires. The purpose of focused Ga$^+$ beam milling here is to generate isolated patterns of magnetic thin film without the need for the time consuming lithography process. However, it is well known that the profile of the Ga$^+$ beam has a long exponential tail in addition to the central Gaussian profile of the ion beam \cite{ref18,ref19}. Therefore, during the sputtering process, the Ga$^+$ irradiation and implantation from the extended tail of the ion beam may affect the ferromagnetic properties of the Py along the edges of the nanowires.  Modification of the ferromagnetic properties including reduction of saturation magnetization of magnetic thin film systems due to ion implantation/irradiation have been reported earlier \cite{ref20,ref21,ref22,ref23,ref24,ref25}. Notably these investigations were performed by ion irradiation/implantation on continuous films and property modifications arose as a direct result of the ion implantation/irradiation the thin film systems. By contrast, in the present investigation, the observed modification occurring along the wire edges of the FIB milled Py nanowires appear to be due to residual irradiation from the extended tail of the Ga$^+$ ion beam. This point will be further discussed in the micromagnetic modeling section C of this paper.

\subsection{Magnetic characterization} \label{II}
\begin{figure}[!hh]
\centering
\includegraphics[width=8cm]{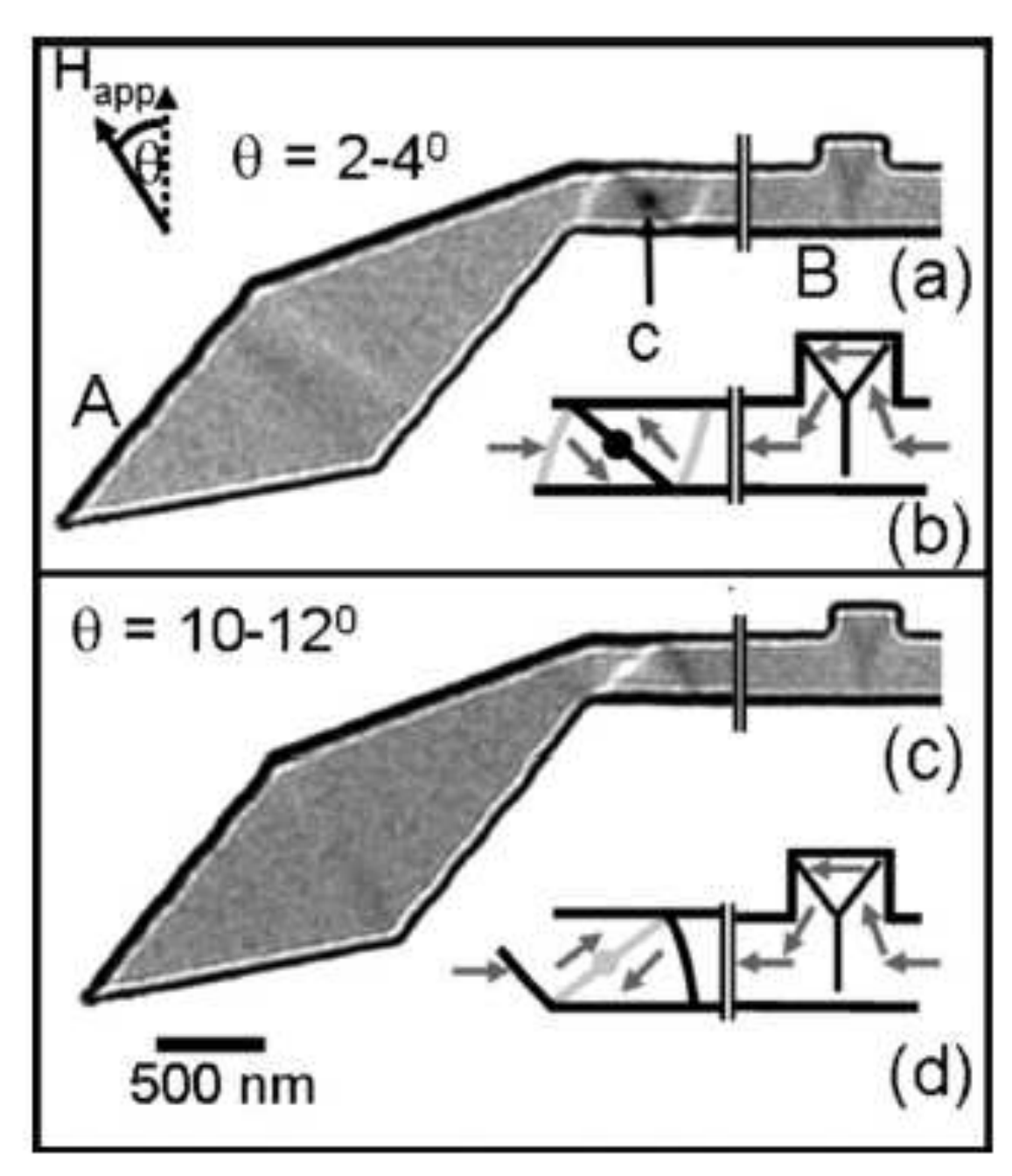}
\caption{Formation of the ccw VDW (a) and cw VDW (c) at the zero field in 320 nm wide EBL patterned nanowires. DWs of different chiralities were formed by applying a magnetic field at different angles counter clockwise from the y-axis and relaxing to zero. For the formation of a ccw VDW, a magnetic field was applied at an angle of 2-4 degrees whereas for a cw VDW formation, the angle of the applied magnetic field was increased to 10-12 degrees. (b and d) are the corresponding schematics of (a and c). 'A' and 'B' indicate the black thick fringe at the upper edge of the diamond shaped pad and lower edge of the straight wire, respectively. 'C' denotes the vortex core.} \label{fig3}
\end{figure}

\emph{In situ} magnetizating experiments were carried out using the Fresnel mode of Lorentz TEM to investigate and compare the magnetic behavior of DWs as they propagate along the EBL and FIB patterned nanowires. Formation of the DWs at the corner position between the pad and the wires were achieved by applying a magnetic field close to the hard axis of the nanowire and then relaxing the field to zero as described in reference \cite{ref26}. For nanowires of 320 nm width and 20 nm thickness, the favored wall structure is a vortex DW (VDW) consistent with the DW structure phase diagram \cite{ref27}. Fresnel images,      Figs. \ref{fig3}(a and c),  show  counter clockwise (ccw) and clockwise (cw) vortex domain wall (VDW) structures, respectively, in EBL patterned nanowires.  

In the Fresnel mode of LTEM, the magnetic DW contrast is clearly visible along with   edge contrast appearing along the wire edges \cite{ref10}. In Fresnel images, Figs. \ref{fig3}(a and c), the appearance of a thick black fringe along the upper edge of the diamond shaped pad (marked by A) and lower edge of the straight wire (marked by B) clearly indicates that these two sections are magnetized in opposite directions  and is consistent with a DW existing between the pad and the wire. The appearance of the vortex, with central section as white or black (indicated by C), depends on the sense of rotation of the magnetization around it. Also visible in Fig.\ref{fig3} is a dark Y-shaped contrast at the anti-notches. From the schematics in Figs. \ref{fig3}(b and d) it can be seen that this results from the magnetization in the nanowire following the edge of the anti-notch resulting in the Y-shaped domain wall structure.
\begin{figure}[!hh]
\centering
\includegraphics[width=7cm]{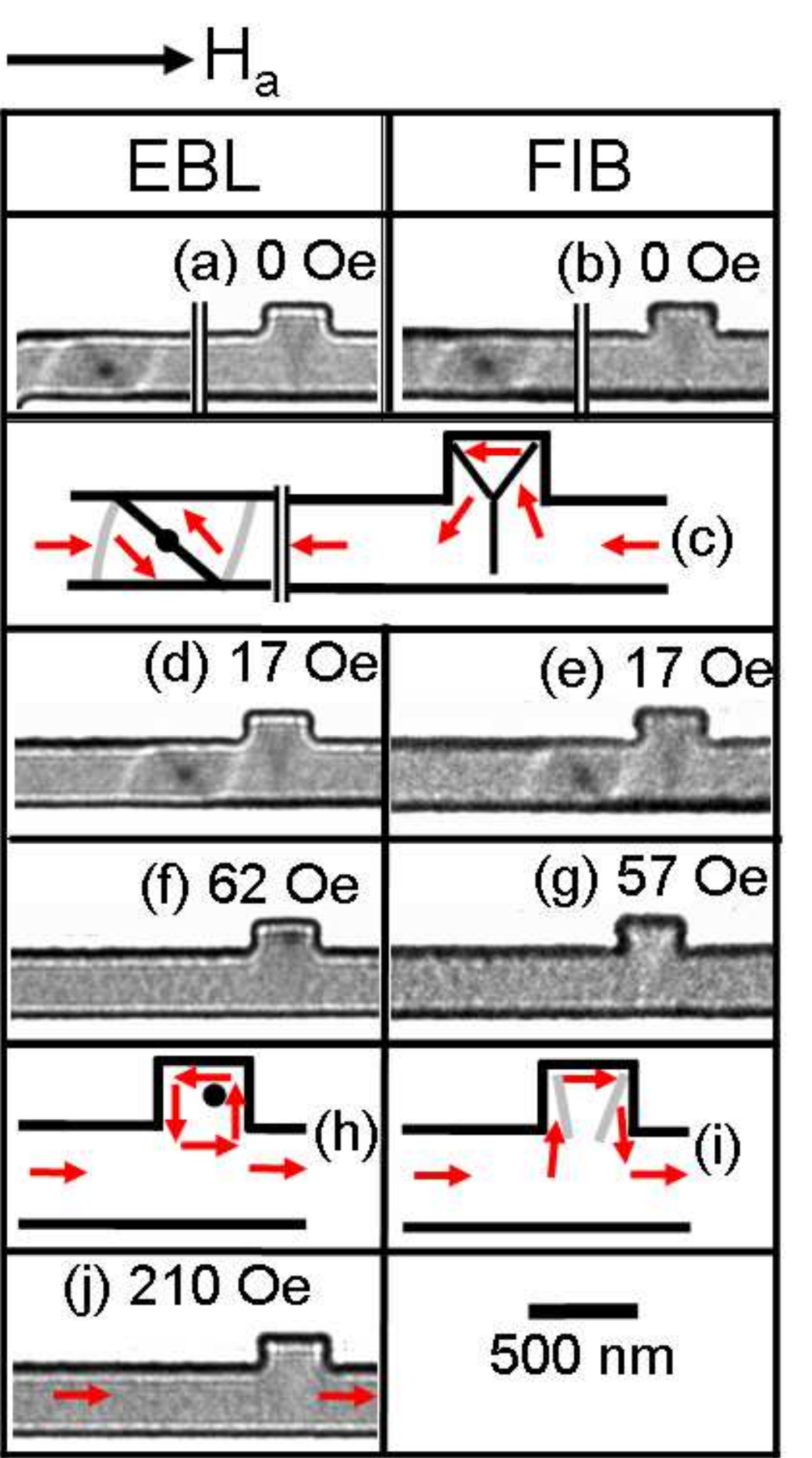}
\caption{Fresnel images of a ccw VDW at zero field in 320 nm wide nanowires patterned by EBL (a) and FIB (b). (d and e) are showing ccw VDW is pinned prior to the anti-notches. State after DW depinning is showing (f and g). Completion of reversal in EBL wire is showing (j). Schematic interpretation of (a) and (b) is (c) and (f and g) are (h and i), respectively.} \label{fig4}
\end{figure}
Following the formation of DWs, a magnetic field was applied along the wire axis to drive the VDWs towards the anti-notches of the nanowires. DW propagation experiments were carried out for both chiralities of VDW in nanowires of both width patterned by EBL and FIB. In terms of reproducibility of the magnetizing behaviour there were 4 identical nanowires for each width and preparation method and each measurement based on 5 repeat cycles of 4 identical nanowires. Therefore the fields quoted in the following results are based on an average of 20 measurements for each event for a given chirality in a wire of a given width for each preparation method. The error comes from the spread of the recorded data. Each set of measurements started from the states shown in Fig. \ref{fig3}  and then followed the sequence propagation, pinning and depinning as the field increased. The state would then be reset and the measurement repeated.  Fresnel images, Figs.  \ref{fig4}(a and b)  show of a ccw VDW at zero field in a 320 nm wide wire patterned by EBL and FIB, respectively. Figure \ref{fig4}(c) is the corresponding schematic showing the magnetization distribution deduced from the images. By applying a magnetic field of 17 Oe, the DWs move and are pinned prior to the anti-notches of both EBL and FIB wires as shown in Figs. \ref{fig4}(d and e), respectively. The ccw VDW was pinned prior to the anti-notch as the magnetization in the leading domain of the DW has a component antiparallel to that in the adjacent anti-notch as can be seen from Fig. \ref{fig4}(c). A further field increase to $\sim {~}$62 Oe results in the DW depinning and propagating through the anti-notch of the EBL nanowire leaving the wire nearly uniformly magnetized, Fig. \ref{fig4}(f). After the depinning of the DW, a vortex of ccw chirality appears inside the anti-notch as a black spot, Fig. \ref{fig4}(f). At an applied magnetic field of $\sim {~}$57 Oe, Fig. \ref{fig4}(g), the magnetization reversal is completed in the FIB nanowire without a vortex forming inside the anti-notch. Here the thicker dark fringe runs along the upper edge of the wire and weak white DW contrast appears inside the anti-notch after the depinning of the DW, which has a similar but opposite configuration to the black Y-shaped domain wall observed earlier but modified by the applied field. The schematic interpretation of Figs. \ref{fig4}(f and g) are (h and i), respectively.  Finally at $\sim {~}$210 Oe, the  vortex structure annihilated from the anti-notch of the EBL patterned nanowire, Fig. \ref{fig4}(j).
\begin{figure}[!hh]
\centering
\includegraphics[width=7cm]{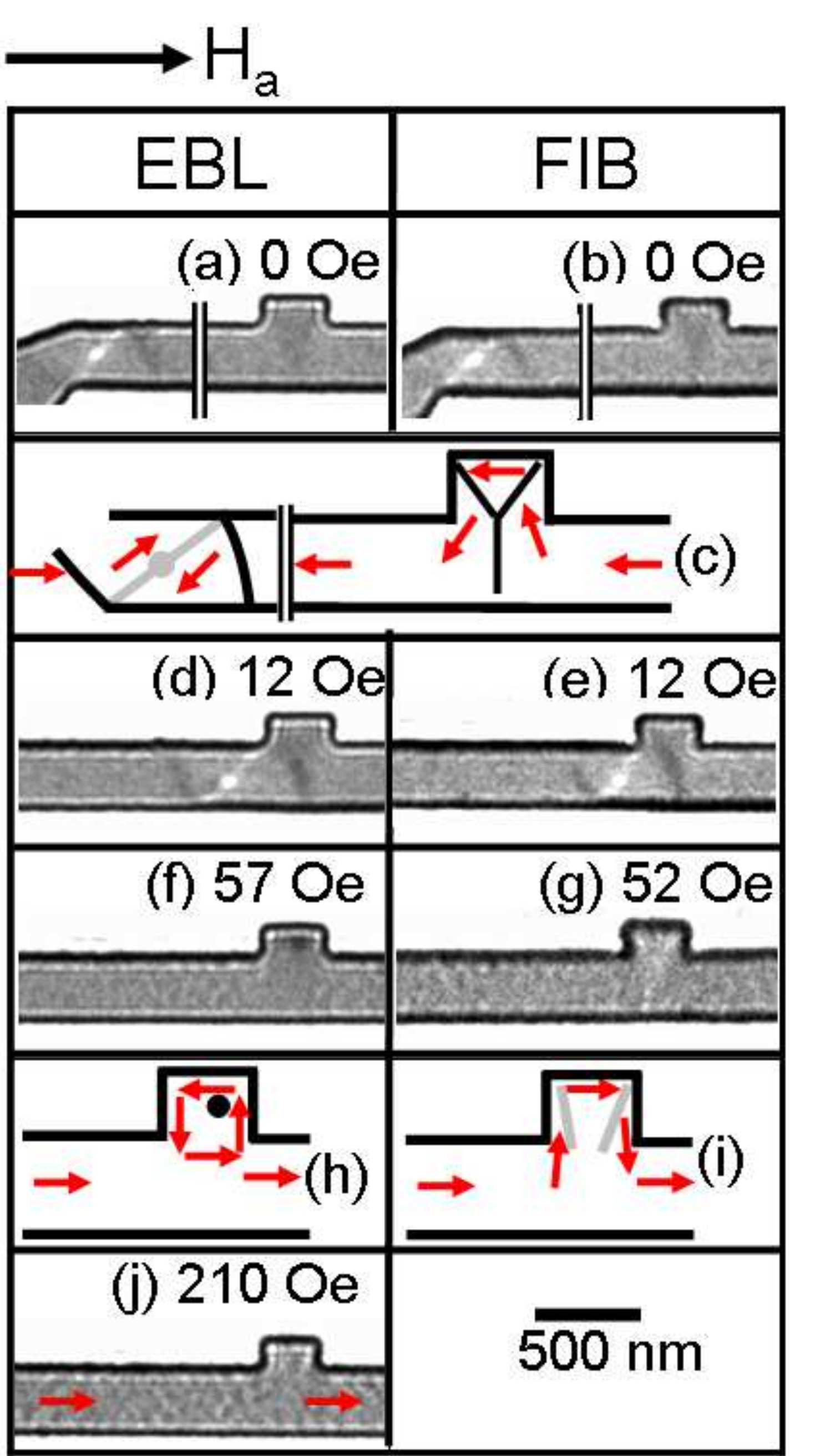}
\caption{Fresnel images (a) and (b) are showing a cw VDW at zero field in 320 nm wide nanowires fabricated by EBL and FIB, respectively. (d and e) are showing part of the cw VDW merged inside the anti-notches. (f and g) are showing state after DW depinning. (j) is showing completion of reversal in EBL wire. (c) is the schematic of (a and b) and (h and i) are the schematics of (f and g), respectively.} \label{fig5}
\end{figure}
Fresnel images, Figs. \ref{fig5}(a and b) show the situation for cw VDWs at zero field in 320 nm wide wires fabricated by EBL and FIB, respectively with a schematic of the magnetization shown in Fig.\ref{fig5}(c). An applied magnetic field of 12 Oe for both cases results in the leading part of the  DW entering the anti-notches as shown in Figs. \ref{fig5}(d,e). As the leading part of the VDWs now have a component parallel to the magnetization at the left side of the anti-notch the VDWs and the Y-shaped domain walls effectively merge. A further field increase to  $\sim {~}$57 Oe results in the DW depinning from the anti-notch of the EBL nanowire, Fig. \ref{fig5}(f). It is noted that after the depinning of the DW, a ccw vortex structure formed inside the anti-notch of the EBL nanowire, as seen by the presence of the black spot in Fig. \ref{fig5}(f). In the FIB milled nanowire, the DW depins at $\sim {~}$52 Oe and weak white DW contrast is formed inside the anti-notch as shown in Fig. \ref{fig5}(g). Figures \ref{fig5}(h and i) are the schematics of (f and g), respectively. Under the application of a magnetic field of $\sim {~}$210 Oe, the  vortex structure is totally removed from the anti-notch of the EBL nanowire, Fig. \ref{fig5}(j). 
 
These results show that DW depinning fields at anti-notches are higher for the ccw vortex domain walls compared to those of the cw vortex domain walls in accordance with previous results on triangular shaped anti-notches in EBL nanowires \cite{ref10}. In the present investigation however the comparison of the EBL and FIB nanowires indicates that DW depinning fields at the anti-notches are lower for the FIB nanowires compared to their equivalent EBL nanowires for both cw and ccw VDWs. Furthermore it is noted that after reversal of the EBL nanowires residual ccw vortex structures are present inside the anti-notch for both chiralities of the incoming DWs. These are not observed in the FIB nanowires after depinning. The presence of these vortices appears to be unrelated to the depinned DWs and it is noted that the same field is required to annihilate the vortex in both cases. Thus the presence of the vortex after depinning does appear to be a residual structure in the anti-notch and not connected to the depinned wall. 
\begin{table}[!h]
\caption{Domain wall depinning field dependency on the fabrication technique and chiralities of the DWs in 320 and 150 nm wide nanowires. Each experimentally measured field is an average from 4 identical nanowires each of which were measured from separate magnetizing cycles. Field values are the mean values of the repeated experiments and error bars represent their standard deviation. Domain wall depinning fields for rectangular (rec.) and sloped (slop.) edge profiles of the nanowires obtained from micromagnetic simulations are also included inside the bottom part of the table.}  
\begin{center}
\begin{tabular}{|l|l|l|l|l|}
 \hline
{Fabrica} & {Wire}  &\multicolumn{2}{c|}{Domain wall}& \multicolumn{1}{c|}{Vortex} \\
{tion} & {width}  &\multicolumn{2}{c|}{depinning field }& \multicolumn{1}{c|}{annihilation} \\
{tech.} & {(nm)}  &\multicolumn{2}{c|}{(Oe)}& \multicolumn{1}{c|}{field (Oe)} \\
\cline{3-4}
 &&ccw&cw &\\
\hline
EBL &320&$62\pm2$&$57\pm2$&$210\pm8$\\
FIB &320&$57\pm2$&$52\pm2$& $----$\\
\hline
EBL &150&$150\pm5$&$118\pm5$&$210\pm8$\\
FIB &150&$129\pm5$&$100\pm8$& $----$\\

\hline
\hline
\hline
Micro &320 (rec. edge) &160&100&300 (ccw/cw VDW)\\
magnetic &320 (slop. edge)&120&100&240 (ccw/cw VDW)\\
\cline{2-5}
\cline{3-4}
simu &150 (rec. edge)&360&200&300 (cw VDW)\\
lation&150 (slop. edge)&320&190&240 (cw VDW)\\
\cline{2-5}
\cline{2-5}
\hline
\end{tabular}
\end{center}
\end{table} \\
Experiments were also performed on EBL and FIB patterned 150 nm wide nanowires which contain anti-notches of width 320 nm and height 140 nm. Like the 320 nm wide wires, a ccw vortex structure appeared inside the anti-notches of the EBL nanowires after the DWs of both ccw and cw chiralities had depinned from the anti-notches. The vortex annihilation field for both chiralities of DWs in 150 nm wide nanowires was again seen to be at $\sim {~}$210 Oe, i.e. the same as for 320 nm wide EBL nanowires.  Again this vortex structure did not appear after the DWs depinned from the anti-notches of the FIB patterned nanowires at this width. Table 1 gives the DW depinning field dependency on the fabrication technique and the chiralities of the DWs along with vortex annihilation field where appropriate. DW depinning fields are higher for the narrower wires compared to the wider wires but also show that FIB nanowires have lower depinning fields for each chirality of DW at both widths. 

The comparative study of the DW depinning in the EBL and FIB nanowires have shown that the depinning fields are comparable in both cases although consistently lower for the FIB nanowires.  It has been seen from another study that nanoelements created by FIB with widths less than 100 nm were not able to be determined as magnetic using LTEM \cite{ref16}. In that case the whole of the magnetic structure appeared to be modified by residual irradiation from the FIB, whereas in this study there is a clearly defined affected edge and non affected internal regions. In the next section we show how micromagnetic simulations can be used to gain an insight into the differing behavior of the EBL and FIB nanostructures based on the possible edge modification in the case of the latter and taking account of previous studies of ion implantation of permalloy. 

\subsection{Micromagnetic simulation} \label{II}

The micromagnetic simulations were carried out using freely available code, object oriented micromagnetic framework (OOMMF) developed by the NIST group at Gaithersburg \cite{ref28}. The parameters used for the simulations were standard for Py: saturation magnetization M$_{s}$ = 8.6 $\times$ 10$^{6}$ \ $A/m$, exchange stiffness constant A = 1.3 $\times$ 10$^{-11}$ \ $J/m$, magnetocrystalline anisotropy K = 0 and damping coefficient $\alpha$ = 0.5.  A cell size of 5 $\times$ 5 $\times$ 5 nm$^3$ was used. Micromagnetic simulations were performed for ccw and cw VDWs in 320 nm and 150 nm wide nanowires which contain 320 nm wide and 140 nm high anti-notches.

\begin{figure}[!hh]
\centering
\includegraphics[width=9cm]{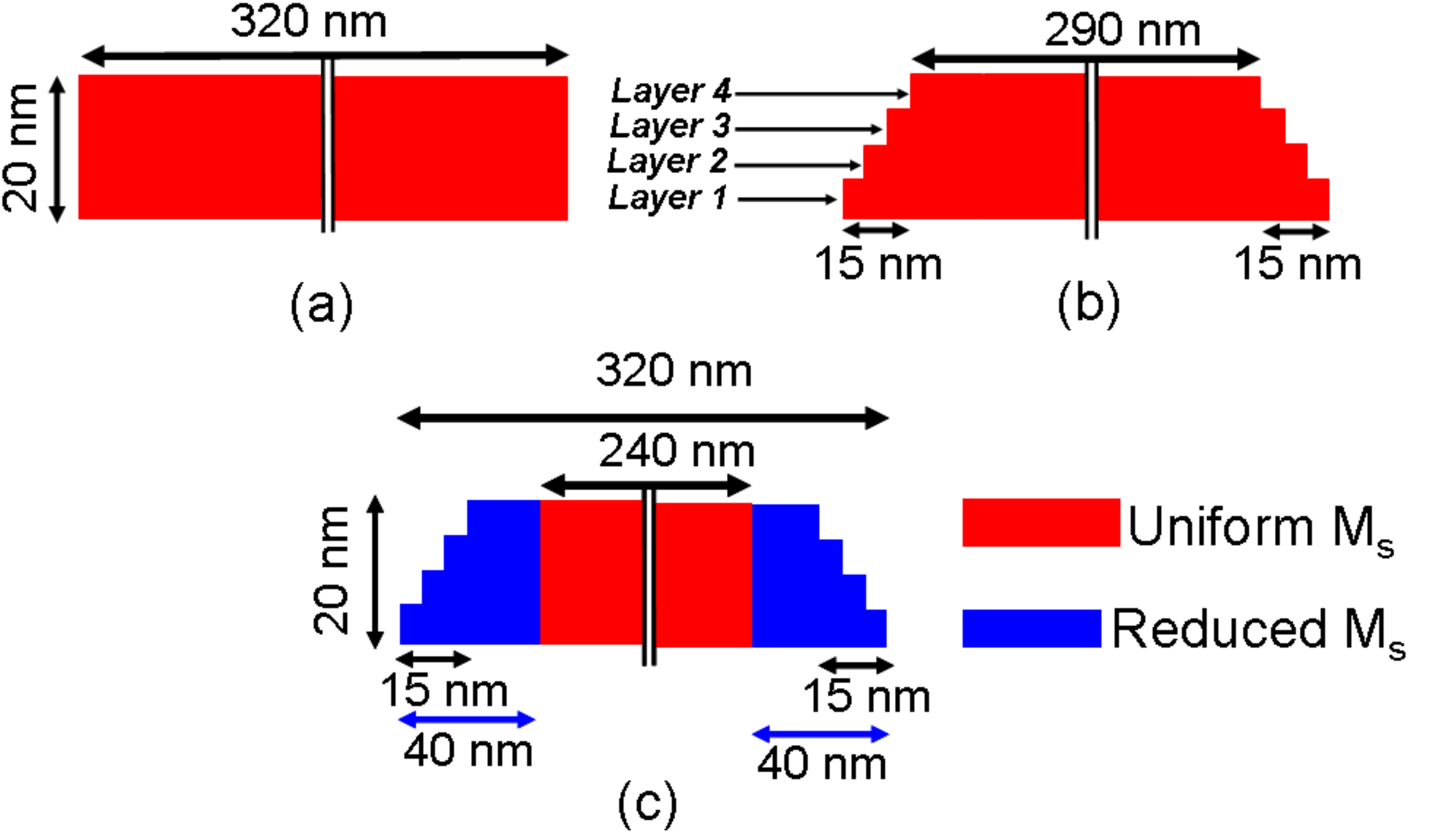}
\caption{Simulations were carried out by varying the edge profiles. Schematic cross-section of the nanowires with rectangular edge (a) and sloped edge (b). (c) M$_{s}$ were reduced up to a distance of 40 nm in the nanowire (b). The width of the nanowires has been reduced for this schematic indicated by vertical bars.} \label{fig6}
\end{figure}

TEM cross-sectional images, Fig. \ref{fig2}(e-h, j) demonstrated a variation in the  edge profile of the  patterned nanowires. Therefore, to conduct simulations initially two different edge profiles were considered. These were shown as schematic cross-section of the wires in Figs. \ref{fig6}(a,b) in which (a) was designed as an ideal structure with a rectangular cross-section and (b) was structured based on the TEM x-sectional image Fig. \ref{fig2} (h) being representative of the experimental edge profile which was sloping rather than vertical. Such an edge profile was modelled by dividing the thickness of the wire into four separate layers, as shown in the schematic \ref{fig6}(b). The bottom layer (layer 1) had the dimensions of the width of the wire with the subsequent layers (layers 2-4) having their width reduced by three cells thus creating a tapered edge profile.

The DW depinning fields obtained from micromagnetic simulations are shown in the lower part of table 1 along with experimentally observed field values. These simulations have confirmed the chirality dependence of the depinning fields i.e. the ccw vortex wall has a higher depinning field than the cw wall. The simulations also show that the wider wires have lower depinning fields than the narrower ones which are consistent with the experimental observations. The effect of the edge structure is evident in that the characteristic fields, i.e. DW depinning and vortex annihilation, are higher for a rectangular edge profile compared to that of the sloped edge profile for all cases considered here. As is the case with previous OOMMF simulations the predicted fields are all much larger than the observed experimental values with simulations being carried out at 0 K \cite{ref7, ref29}, however the trends of the higher characteristic fields for rectangular edges narrower wires are reproduced. Whilst the sloped edge structure accounts for lower characteristic fields like the experimental FIB structures it does still retain the residual vortex structure after DW depinning. Although in the case of the ccw VDWs in the 150 nm wide wires the simulated depinning fields are greater than the vortex annihilation field seen in the cw DW depinning. Figs. 7(a-d) shows the sequence of Fresnel images simulated for a ccw DW moving through a nanowire, which shows good agreement with the process for the EBL fabricated nanowire in Fig. \ref{fig4}. 

\begin{figure}[!hh]
\centering
\includegraphics[width=9cm]{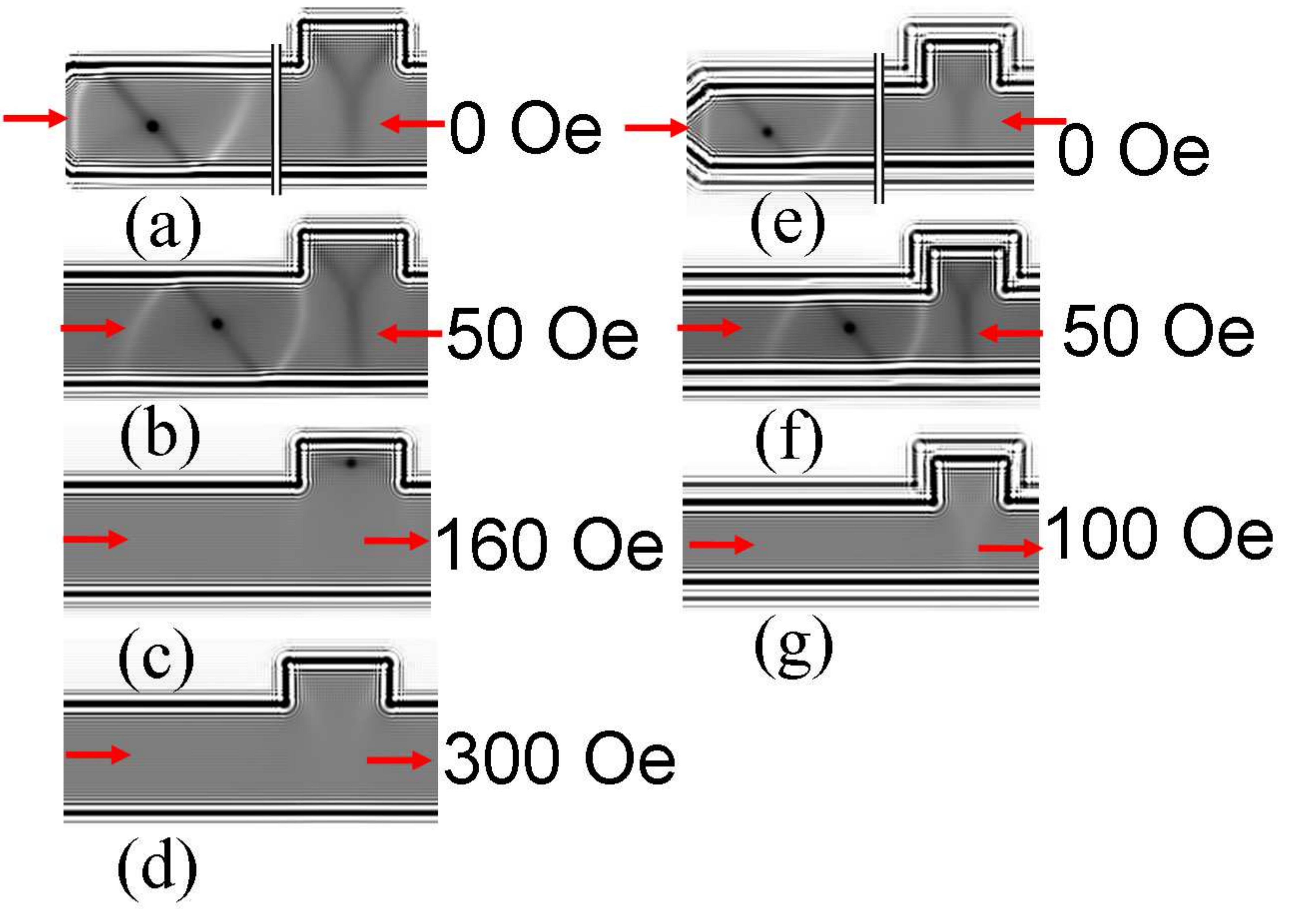}
\caption{(Color online) Simulated Fresnel images (a-d) show the magnetization reversal process in 320 nm wide nanowire (sloped eedge) with uniform M$_{s}$ throughout the wire. DW has moved through the anti-notch and a vortex structure is formed inside the anti-notch, as shown in (c). Simulated Fresnel images (e-g) are showing the reversal process of the nanowire of same geometry but with M$_{s}$ is 50\% along the wire edge. DW has moved through the anti-notch without the presence of the vortex structure, as shown in (g).} \label{fig7}
\end{figure}

These initial simulations have shown that the characteristic DW depinning fields are reduced for a sloped edge nanowire; however, they do not explain why  residual vortices are not seen in the anti notches after DW depinning. As mentioned earlier, it was assumed that during the milling process of the nanowires by focused Ga$^+$  beam, the wire edges were  affected by Ga$^+$ irradiation due to the extended tail of the Ga$^+$ beam. Previously studies have look at the effect on films irradiated homogeneously with 30 keV Ga$^+$, for example investigation \cite{ref20} was carried out using Magneto-optic Kerr effect measurements at room temperature on Ni$_{80}$Fe$_{20}$ (15.5nm)/Ni$_{80}$Cr$_{20}$(9.0nm). This study demonstrated coercivity changes in a 100 $\ \mu m^2$  area even for mildly dosed (8.0 x 10$^{14}$ ions/cm$^2$) samples that did not receive more than 1 at.\% of Ga.  However, a dramatic reduction in the Curie temperature was observed for doses of 1 x 10$^{16}$ ions/cm$^2$ for direct Ga$^+$ implantation. It may also be noted here that doping Ni$_{80}$Fe$_{20}$  alloys with other metal such as Cr also causes the Curie temperature and saturation magnetization to be reduced, and that at 8 at\%  Cr the alloy becomes paramagnetic at room temperature \cite{ref30}. Therefore, to clarify the experimentally observed two-step reversal process in the case of EBL patterned nanowires and one step process for FIB nanowires, micromagnetic simulations were also carried out by varying the saturation magnetization (M$_{s}$) of Py along the nanowire edges. 

To observe the influence of the reduction of M$_{s}$ along the wire edge, a similar wire structure for Fig. \ref{fig6}(b)  but with a reduced M$_{s}$ up to a distance of 40 nm from the wire edge was considered as shown schematically in the blue marked region in Fig. \ref{fig6}(c). As was mentioned earlier, in the plan view bright field image Fig. \ref{fig2}(d), most of the large size grains were observed up to a distance of  40 nm from the wire edge of the FIB milled nanowire. This was the basis for reducing M$_{s}$ in simulations  up to this distance of the wire edges.

\begin{figure}[!hh]
\centering
\includegraphics[width=9cm]{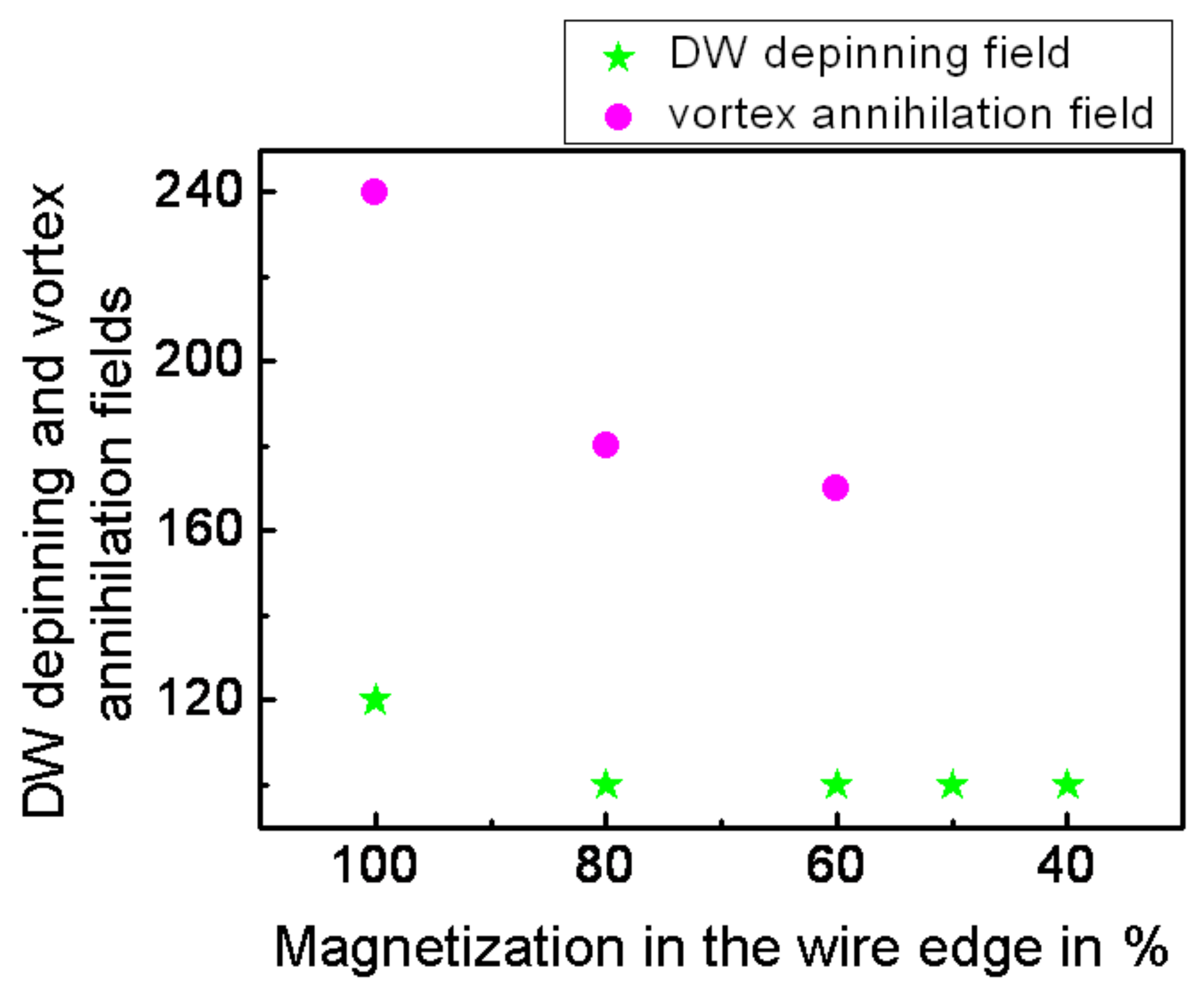}
\caption{(Color online) DW depinning and vortex annihilation fields as a function of M$_{s}$ along the wire edge. The simulations were carried out for a ccw VDW in a 320 nm wide and 20 nm thick wire as shown in the schematic cros-section \ref{fig6}(c).} \label{fig8}
\end{figure}

The simulations were performed by varying $M_{s}$ up to a distance of 40 nm from the edge along the edges for a ccw VDW in a 320 nm wide and 20 nm thick wire as shown in the schematic cross-section Fig. \ref{fig6}(c). The $M_{s}$ value in the central part of the wire remained at 100\%. The dimension of the anti-notch was fixed as before, i.e. 320 nm wide and 140 nm high.  $M_{s}$ values along the wire edge were reduced from 100\% down to 40\% and results were plotted in Fig. \ref{fig8}. This figure shows that if $M_{s}$ is reduced along the wire edges the DW depinning field decreases by around 20\% compared to the unmodified wire. Additionally it can be seen that the vortex annihilation field inside the anti-notch also reduces with reduced $M_{s}$ along the wire edge.  However when $M_{s}$ is reduced to 50 \% along the wire edge, the ccw vortex structure is no longer formed inside the anti-notch of the wire. To illustrate this the reversal process for a 50 \% $M_{s}$ along the wire edge of a 320 nm wide wire  is shown in Figs. \ref{fig7}(e-g). This suggests that the experimental results, Figs.\ref{fig4} and \ref{fig5}(b,e and g) for FIB patterned nanowires may be consistent with reduced edge $M_{s}$, due to Ga ion implantation.

\section{Conclusions} \label{II}
In the present investigation, high quality nanostructures were fabricated by using EBL and FIB techniques and characterized by TEM imaging.  Bright field images revealed significant grain growth along the wire edges of the FIB patterned nanowires. However, TEM x-sectional images have confirmed that well defined edge profile result for nanowires patterned using both EBL and FIB techniques. Magnetizing experiments using the Fresnel mode of Lorentz TEM revealed differences in the magnetic behavior between EBL and FIB patterned nanowires, notably small but reproducible differences in the depinning fields and the absence of residual vortices in the anti-notches of the FIB nanowires. Evidence from micromagnetic simulations suggest that this different reversal process between the nanowires fabricated by the FIB techniques may be associated with the reduction of M$_{s}$ along the wire edges. Such a reduction of M$_{s}$ along the wire edges reduces the DW depinning fields and is consistent with the experimentally observed magnetization behavior of the FIB milled nanowires. Indeed this point is consistent with a study of nanoelements of with dimensions $<$ 100 nm in which the whole element appeared to be affected by the residual irradiation from the FIB \cite{ref16}. The difference in the reversal process between EBL and FIB patterned nanowires are also demonstrated by the formation of a ccw vortex structure inside the anti-notches of the EBL nanowires after the depinning of DWs in contrast to the FIB nanowires where no such vortex structures formed. The presence of these vortices could potentially be very problematic for nanowire applications where series of domain walls pass through wires in proposed memory and logic applications \cite{ref1,ref2}. Therefore for the wire dimensions considered here FIB may actually be beneficial as a fabrication method. However it must be noted that the edge modification induced by the FIB method does have implications particularly if smaller dimensions are to be fabricated and this certainly needs to be carefully considered when using this method.
\\

\section{Acknowledgements}
The author (M A Basith) would like to acknowledge sincerely University of Glasgow and ORSAS for providing financial support.

*Email address: m.basith@physics.gla.ac.uk

\end{document}